\documentclass[a4paper,preprint,aps,prb,showpacs]{revtex4-1}
\usepackage{graphicx,amsmath,amssymb}

\renewcommand{\[}{\begin{equation}}
\renewcommand{\]}{\end{equation}}
\def\bea{\begin{eqnarray}}
\def\eea{\end{eqnarray}}
\def\nn{\nonumber\\}
\newcommand{\equ}[1]{Eq.~(\ref{#1})}
\newcommand{\eqs}[2]{Eqs.~(\ref{#1}) and (\ref{#2})}

\def\r{{\bf r}}
\def\k{{\bf k}}
\def\C{{\mathfrak C}}
\newcommand{\lt}{\langle r_\alpha r_\beta \rangle_{\rm c}}
\newcommand{\intk}{\int_{\rm BZ} \!\!\!\!\! d{\bf k}}

\newcommand{\emi}[1]{{\rm e}^{-i #1}}
\renewcommand{\r}{{\bf r}}

\newcommand{\nc}{N_{\rm c}}
\def\bra#1{\langle#1\vert}
\def\ket#1{\vert#1\rangle}
\def\ev#1{\langle#1\rangle}
\def\me#1#2#3{\langle#1| \, #2 \, |#3\rangle}
\begin{document}
\title{Mapping topological order in coordinate space}

\author{Raffaello Bianco and Raffaele Resta}

\affiliation{Dipartimento di Fisica, Universit\`a di Trieste, Italy,
and DEMOCRITOS National Simulation Center, IOM-CNR, Trieste, Italy}


\begin{abstract}
The organization of the electrons in the  ground state is classified by means of topological invariants, defined as global properties of the wavefunction. Here we address the Chern number of a two-dimensional insulator and 
we show that the corresponding topological order can be mapped  by means of a ``topological marker'', defined in  $\r$-space, and which may vary in different regions of the same sample. Notably, this applies equally well to periodic and open boundary conditions. Simulations over a model Hamiltonian validate our theory.
\end{abstract}

\date{}


\maketitle \bigskip\bigskip

Topological insulators are sharply distinguished from normal ones by the manner in which the electronic ground state is topologically ``twisted'' or ``knotted'' in $\k$-space.\cite{Moore10,Hasan10} But topological order must reflect a peculiar organization of the electrons even when the concept of $\k$-space does not apply, such as for inhomogeneous systems, as well as for finite systems within open boundary conditions. We address here the archetypical topological invariant, namely the first Chern number  $C$, defined for a many-electron system in two dimensions (2d), and we show that the corresponding topological order also bears a very clear signature in $\r$-space. We introduce a  ``topological marker'', which may vary in different regions of the same sample, and  we validate our expression by means of simulations on a model Hamiltonian, performed on finite samples within {\it open} boundary conditions. Our test cases include crystalline as well as disordered samples, and
  heterojunctions.

For a lattice-periodical system of independent electrons the Chern number 
(a.k.a. TKNN invariant\cite{Thouless82}) $C$ is expressed as a 2d Brillouin-zone integral. For a disordered and macroscopically homogeneous system  $C$ has a known expression in a supercell framework,\cite{Yang96,rap135} also formulated in $\k$-space. 
The concept of $\k$-space is rooted in the periodic boundary conditions (or generalizations thereof), while instead  our topological marker samples the electronic ground state {\it locally}. The choice of boundary conditions becomes irrelevant in the limit of a large sample.

For a system of independent electrons, within either periodic or open boundary conditions, the ground state is uniquely determined by the one-particle density matrix, a.k.a. ground-state projector $P(\r,\r')$; it is a ``nearsighted''\cite{Kohn96,rap132,rap_a31} operator, exponentially decreasing with $|\r - \r'|$ in insulators even when $C \neq 0$.\cite{Thonhauser06} Our major result is expressing the topological marker in terms of $P$ directly, 
\eqs{tilde}{local} below, where the one-particle orbitals do not appear.

Let $u_{n\k}(\r) = \emi{\k \cdot \r} \psi_{n\k}(\r)$ be the periodic part of the Bloch orbitals, normalized in the unit cell of area $A_{\rm c}$. The standard expression of the Chern invariant in a 2d lattice-periodical insulator is\cite{sign}  \[ C  = - \frac{1}{\pi}  \mbox{Im }  \sum_{n=1}^{\nc}   \intk  \ev{\frac{\partial}{\partial k_x} u_{n\k} | \frac{\partial}{\partial k_y} u_{n\k}} \; ; \label{chern} \] 
we assume single occupancy (a.k.a. ``spinless electrons'') throughout.
In \equ{chern}  $\nc$ is the number of occupied bands, and the integral is over the Brillouin zone;  $C$ is guaranteed to be an integer and is gauge invariant, i.e. invariant either by unitary transformations of the occupied orbitals among themselves, or by a change of the magnetic gauge. It characterizes therefore the many-electron ground state, not the Hamiltonian. 

We start inserting a complete set of states into \equ{chern}  \[ C =  - \frac{1}{\pi} \mbox{Im } \sum_{n=1}^{\nc}  \sum_{n'=\nc+1}^{\infty}  \intk \ev{\frac{\partial}{\partial k_x} u_{n\k} | u_{n'\k} } \ev{ u_{n'\k}
 | \frac{\partial}{\partial k_y} u_{n\k}} ,  \] where the missing terms are real. Then, by some manipulations which are standard in linear-response theory,\cite{Baroni01} we have \[ \ev{ u_{n'\k}|\nabla_{\k} u_{n\k}}  = -i \me{\psi_{n'\k}}{\r}{\psi_{n\k}} , \qquad n\neq n' . \label{def} \] We stress that, while the position operator $\r$ is ill-defined within periodic boundary conditions,\cite{rap100} its off-diagonal elements over the Hamiltonian eigenstates are well defined; more accurately, \equ{def} should be interpreted as a {\it definition} of such elements. Then
 \bea C &=& -
 \frac{1}{\pi}  \mbox{Im }  \sum_{n=1}^{\nc}  \sum_{n'=\nc+1}^{\infty} \intk  \me{\psi_{n\k}}{x}{\psi_{n'\k}}  \me{\psi_{n'\k}}{y}{\psi_{n\k}}  \nn &=& -
 \frac{1}{\pi}  \frac{A_{\rm c}}{(2\pi)^2}  \mbox{Im } \sum_{n=1}^{\nc}  \sum_{n'=\nc+1}^{\infty} \intk \intk' 
\me{\psi_{n\k}}{x}{\psi_{n'\k'}}  \me{\psi_{n'\k'}}{y}{\psi_{n\k}} , \label{p1} \eea where the second line owes to the fact that the matrix elements vanish for $\k \neq \k'$. Next we recognize the ground-state projector $P$ and its complement $Q=1-P$ 
\bea P &=&  \frac{A_{\rm c}}{(2\pi)^2}\sum_{n=1}^{\nc} \intk \ket{\psi_{n\k}}\bra{\psi_{n\k}} , \nn Q &=& \frac{A_{\rm c}}{(2\pi)^2}\sum_{n'=\nc+1}^\infty \intk' \ket{\psi_{n'\k'}}\bra{\psi_{n'\k'}} .  \eea
\equ{p1} becomes then the trace over a crystal cell of a real-space operator:
\bea C &=& -
 \frac{1}{\pi}  \frac{(2\pi)^2}{A_{\rm c}} \mbox{Im } \mbox{tr}_{\rm cell} \{ PxQy \} \nn &=&   \frac{4 \pi }{A_{\rm c}} \mbox{Im }  \mbox{tr}_{\rm cell} \{ PxPy \}  , \label{trc} \eea where the symmetry of the missing term yields the second line. By exploiting the general properties of projectors and of the trace, \equ{trc} can be recast in several equivalent ways. For lattice models, a similar real-space formula has been demonstrated in 2006 by Kitaev;\cite{Kitaev06}
to the best of our knowledge, we are providing the first proof which does not rely on lattice models and generalizes Kitaev's result to realistic implementations.

Subsequent work adopting Kitaev's formula was invariably  rooted in $\k$-space within a toroidal geometry, for a system without boundaries, and was based on traces.\cite{Prodan09,Prodan10,Prodan10b} Finite systems within open boundary conditions look problematic. In fact, if we replace the trace over the cell with the trace over the whole sample, the identity \[  \mbox{Im }  \mbox{tr} \{ PxPy \} = \frac{1}{2i} \mbox{tr}\{ \, [ PxP , PyP ] \, \}  \label{commu}  \] guarantees a zero result, whenever $P$ projects over a finite-dimensional manifold. This confirms that the global topology is trivial within open boundary conditions, and also hints that traces must be avoided when addressing finite and/or inhomogeneous samples.

At variance with previous work based on Kitaev's formula, we propose here to directly address the commutator in \equ{commu} before taking the trace. Let $\tilde{X}$  be the projected $x$-coordinate  \[ \tilde{X}(\r,\r') = \int d \r'' \; P(\r,\r'') x'' P(\r'',\r') , \label{tilde} \] and similarly $\tilde{Y}$;  we then identify the topological marker with the {\it local} Chern number as\cite{sign} \[ \C(\r) = - 2 \pi i \int d\r' \; [ \, \tilde{X}(\r,\r') \tilde{Y}(\r',\r) - \tilde{Y}(\r,\r') \tilde{X}(\r',\r) \, ] . \label{local} \] Our definition holds within both periodic and open boundary conditions; given the shortsightedness of $P$, in a region of crystalline periodicity the cell average of $\C(\r)$ coincides with the Chern number $C$ owing to \equ{trc}. We expect the dimensionless function $\C(\r)$ to fluctuate over microscopic dimensions; in the nonperiodic case,
the cell average has to be replaced with the macroscopic average, defined as in electrostatics (see e.g. Jackson\cite{Jackson}).

The gauge invariance of $\C(\r)$ as defined in \equ{local} deserves a comment. The ground-state projector $P$ is invariant by unitary transformations of the occupied orbitals among themselves, but {\it not} by a change of the magnetic gauge. However, the unitary operator which transforms $P$ is local in coordinate space, thus ensuring gauge invariance of $\C(\r)$.
 
\begin{figure}
\centering
\includegraphics[width=.9\columnwidth]{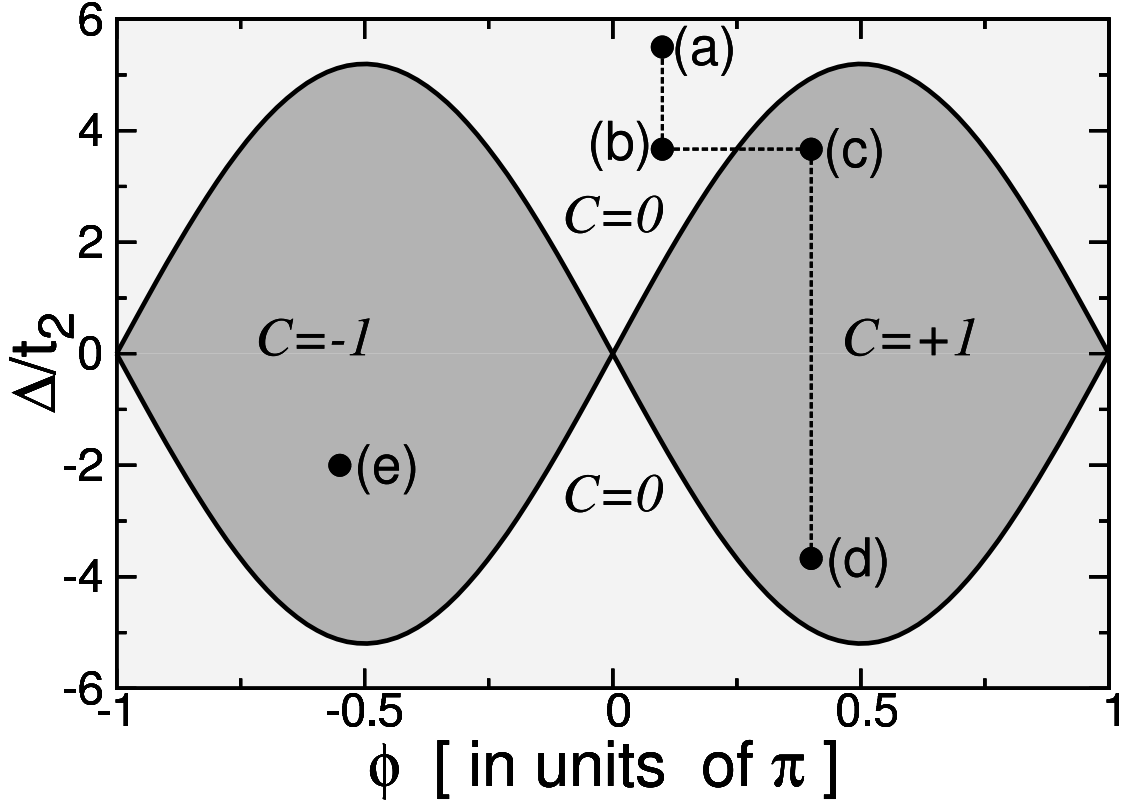}
\caption{Chern number of the bottom band of the Haldane model as a function of the parameters $\phi$ and $\Delta/t_2$ ($t_1=1, t_2=1/3$). The points marked with letters (a-e) in this phase diagram are relevant for the subsequent discussion and figures. In order to avoid special features the $\phi$ parameter is {\it not} a multiple of $\pi/4$.}
\label{fig:diagram} \end{figure}

\begin{figure}
\centering
\includegraphics[width=.9\columnwidth]{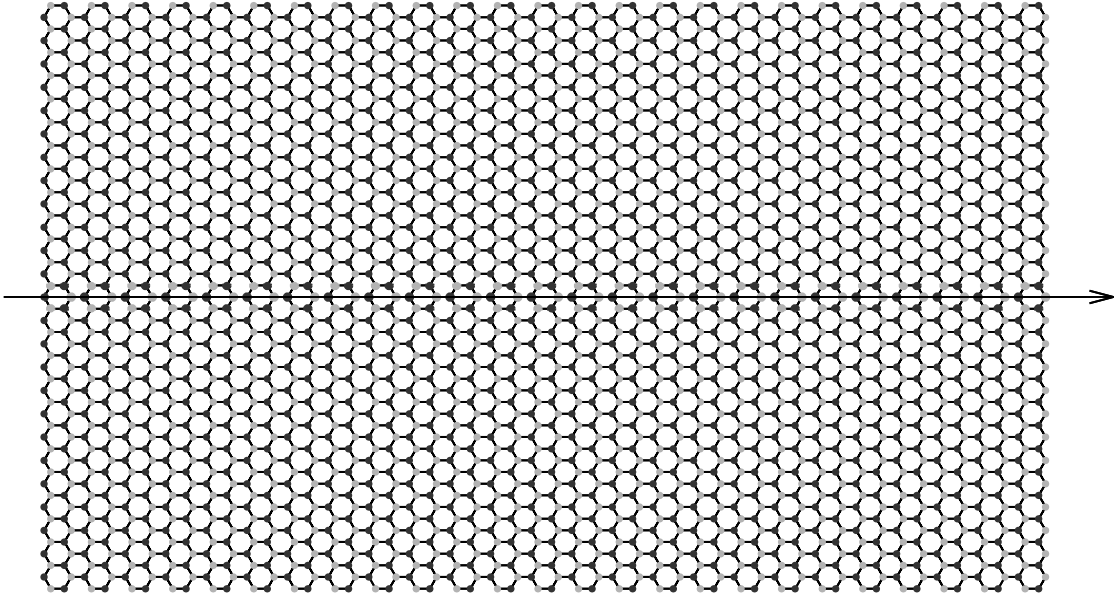}
\caption{A typical flake, with 2550  sites, showing the honeycomb lattice of the Haldane model\protect\cite{Haldane88}. The 50 sites on the horizontal line  will be used in all the subsequent one-dimensional plots. Black and grey circles indicate nonequivalent sites (with onsite energies $\pm \Delta$)}
\label{fig:flake} \end{figure}

We validate our formal findings by performing simulations on the Haldane model Hamiltonian;\cite{Haldane88} it is
comprised of a 2d  honeycomb lattice with two tight-binding sites per primitive
cell with site energies $\pm \Delta$, real first-neighbor hoppings $t_1$, and
complex second-neighbor hoppings $t_2e^{\pm i\phi}$. As a function of the parameters, this 2d model system may have either $C = 0$ or $C = \pm 1$, according to the phase diagram shown in Fig. \ref{fig:diagram}. This model has been previously used in several simulations, providing invaluable insight into orbital magnetization\cite{rap128,rap130,rap135} as well as into nontrivial topological features of the electronic wavefunction.\cite{Haldane88,rap135,Thonhauser06,Hao08,Coh09}  At half filling the system is insulating, except when $\Delta = t_2  \sin \phi =0$. In the present work we study, within open boundary conditions, finite flakes of rectangular shape cut from the bulk, as shown in Fig. \ref{fig:flake}. We have addressed homogenous samples where the Hamiltonian is chosen from various points of the phase diagram, Fig. \ref{fig:diagram}, as well as disordered and inhomogeneous samples.

\begin{figure}
\centering
\includegraphics[width=0.9\columnwidth]{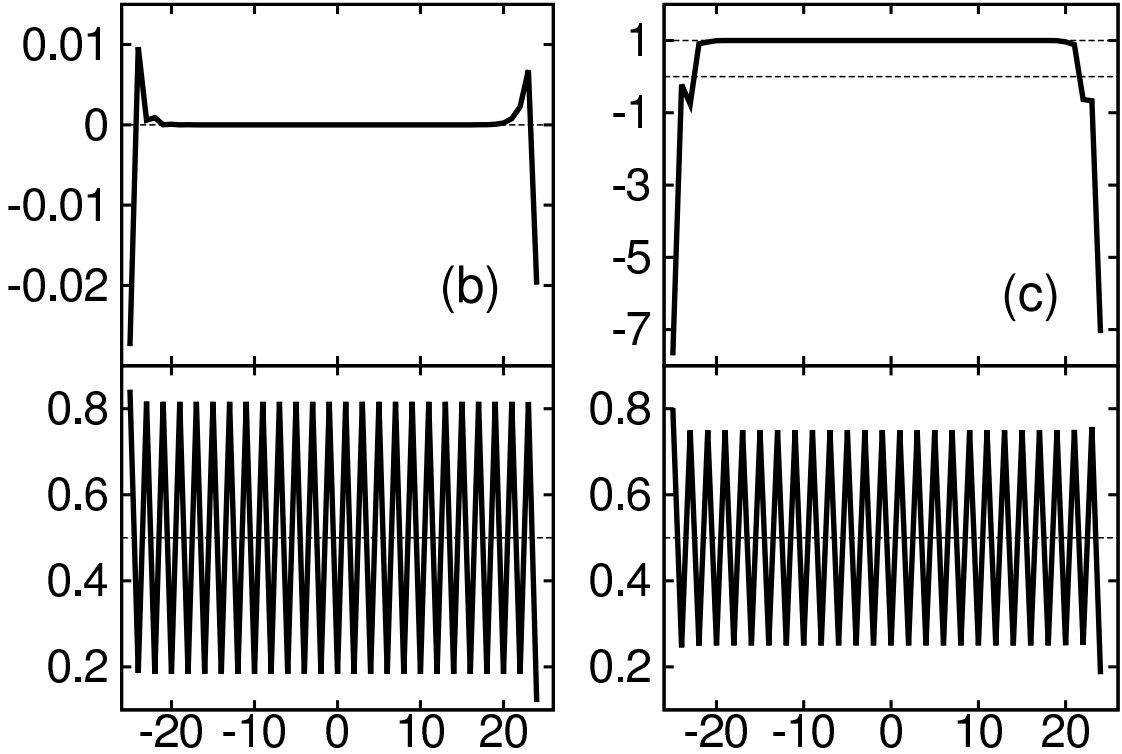}
\caption{Local Chern number (top) and site occupancy (bottom), for the 50 sites along the line shown in Fig. \protect\ref{fig:flake}. Left panel: point (b) in the phase diagram, Fig. \protect\ref{fig:diagram}. Right panel: point (c). Notice the different scales.}
\label{fig:crys} \end{figure}

Two typical plots for crystalline samples are shown in Fig. \ref{fig:crys}, where we have chosen the two points (b) and (c) in Fig. \protect\ref{fig:diagram}, with $C=0$ and $C=1$, respectively. The plots confirm that the local Chern numbers $\C(i)$ are equal to either 0 or 1 (as expected) in the bulk of the sample, while they deviate in the boundary region. In both cases the negative values compensate for the positive ones, given that the sum of the $\C(i)$ over the whole sample vanishes. This compensation is most interesting when $C=1$ (right panel).
A size analysis shows that the minimum negative $\C(i)$ value scales like $L$ (linear dimension of the sample): the reason is that  the number of bulk sites scales as $L^2$, while the perimeter scales as $L$. 

We have studied both polar ($\Delta \neq 0$) and nonpolar ($\Delta = 0$) cases. While in the latter case the two sites are equivalent, they are no longer so in the former case. This  clearly appears in the  site occupancies, also shown Fig. \ref{fig:crys}. What is surprising, is that the corresponding $\C(i)$ values do not show any site alternance, while we expect only their cell (or macroscopic) average to be equal to one. We conjecture this to be due to extra symmetry present in the Haldane model Hamiltonian, actually broken in disordered samples, discussed below (see Fig. 5).

\begin{figure}
\centering
\includegraphics[width=.75\columnwidth]{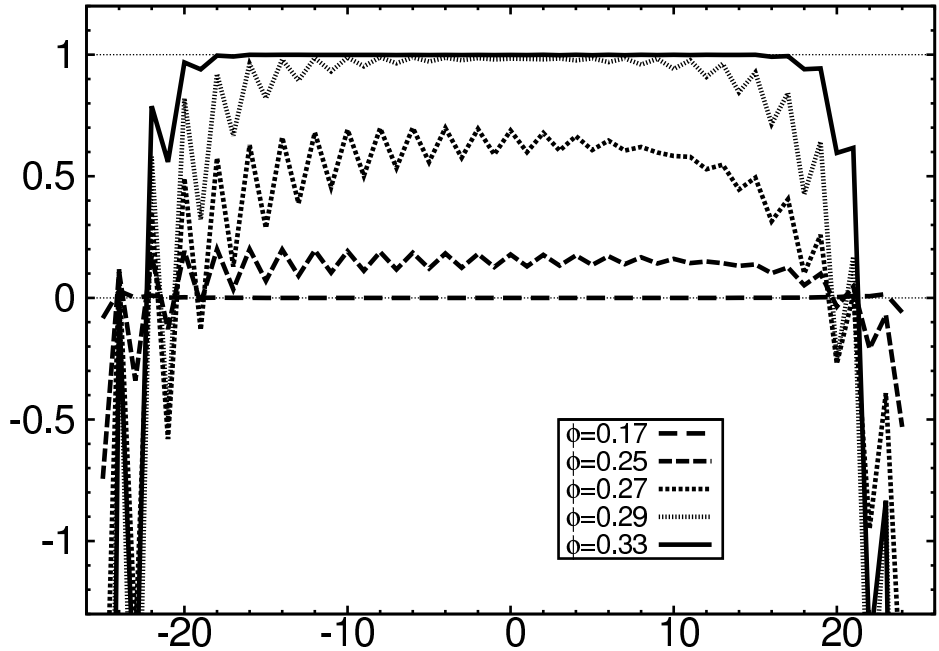}
\caption{Local Chern number for a few points on the line $\Delta/t_2 = 3.67$, i.e. on the (b-c) segment in Fig. \protect\ref{fig:diagram}, close to the transition from $C=0$ to $C=1$. The exact transition occurs at $\phi/\pi=0.25$;
our five plots correspond (bottom to top) to $\phi/\pi=0.17,0.25,0.27,0.29,0.33.$
}
\label{fig:trans} \end{figure}

We have also investigated a few points in the phase diagram close to the transition between $C=0$ and $C=1$ at fixed $\Delta/t_2 = 3.67$ and various $\phi$ values. Given the finite size of the system the transition cannot be sharp. The exact transition for an infinite system occurs at $\phi/\pi=0.25$; our results show that in the bulk of the sample the local Chern number is zero up to $\phi/\pi \simeq 0.17$ and one from $\phi/\pi\simeq 0.29$ onwards. At intermediate values the boundary region broadens considerably and indeed invades the whole sample: this is shown in Fig.~\ref{fig:trans}.

\begin{figure}
\centering
\includegraphics[width=0.9\columnwidth]{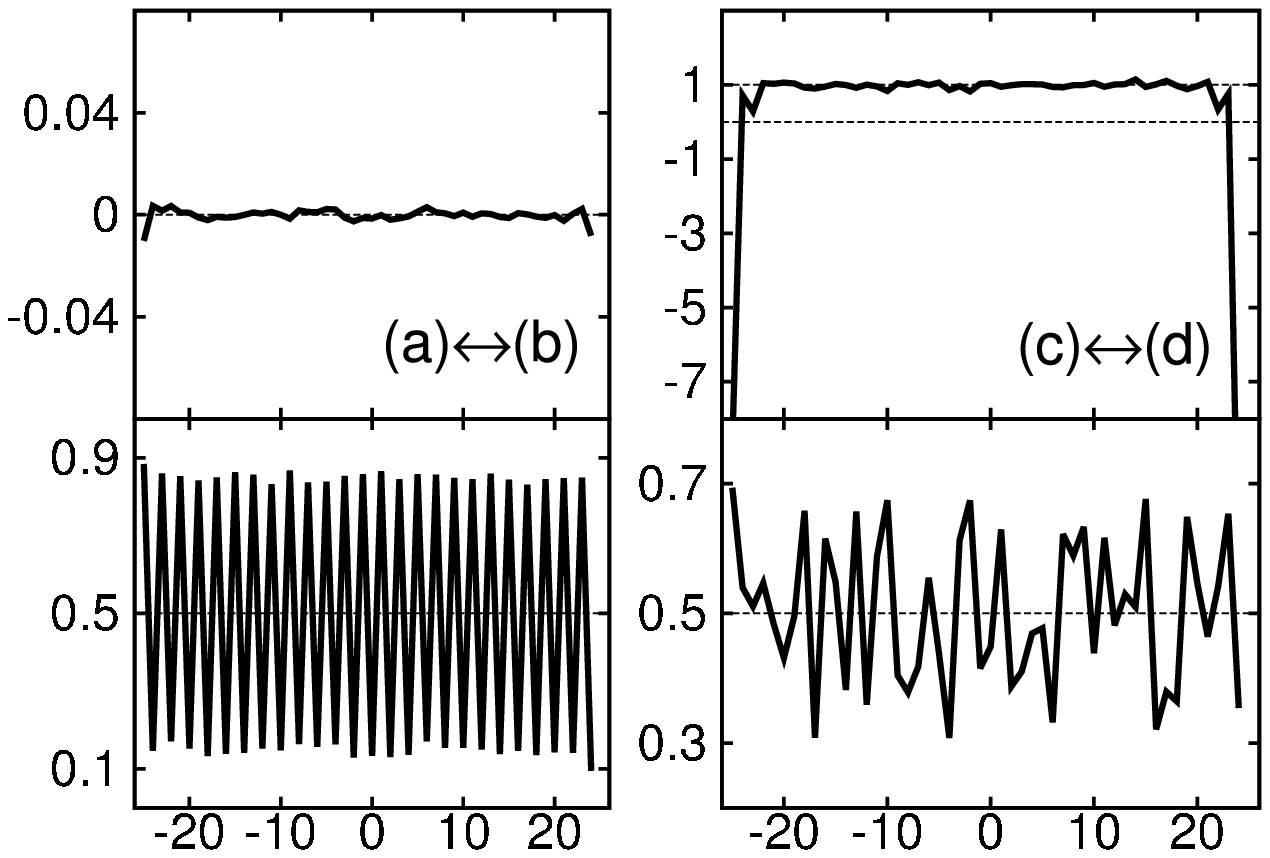}
\caption{Local Chern number (top) and site occupancy (bottom), for disordered systems (see text). Left panel: disordered system along the line (a-b) in the phase diagram, Fig. \protect\ref{fig:diagram}. Right panel: line (c-d). Notice the different scales.}
\label{fig:dis} \end{figure}

Typical results for disordered---and macroscopically homogenous---samples are shown in Fig. \ref{fig:dis}. In the left panel the sign of $\Delta$ alternates between the two sublattices, while its modulus is chosen at random (with uniform distribution) in the (a-b) segment of  Fig. \protect\ref{fig:diagram}. In the right panel the value of $\Delta$ is chosen at random in the (c-d) segment.
It appears clearly that the local Chern numbers $\C(i)$ in the bulk of the sample oscillate around a macroscopic average $C=0$ (left panel) and $C=1$ (right panel).

\begin{figure}
\centering
\includegraphics[width=0.9\columnwidth]{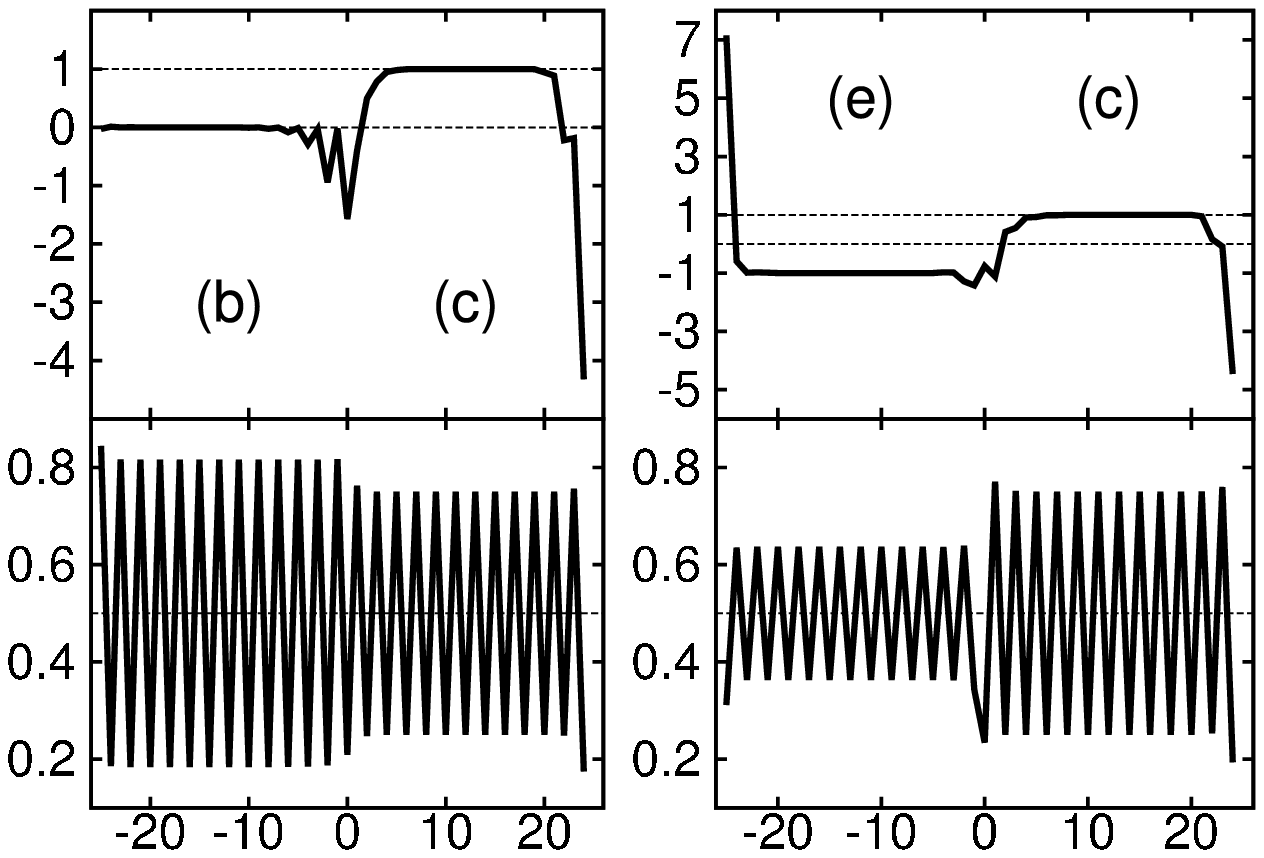}
\caption{Local Chern number (top) and site occupancy (bottom), across heterojunctions. Left panel: Hamiltonian parameters as in (a) and in (b) for left and  the right halves of the sample, respectively. Right panel:  parameters as in (e) and in (c) for left and  the right halves of the sample.}
\label{fig:hetero} \end{figure}

Next we show in Fig. \ref{fig:hetero} our topological marker across an heterojunction between regions of different topological order, in two typical cases: a normal insulator joined to a $C=1$ insulator, and a junction where $C$ changes sign. In both cases the marker maps very perspicuously the actual topological order in the two bulklike regions, while it oscillates at the interface and at the sample boundary.
The virtue of our $\r$-space approach is clearly demonstrated; the conventional $\k$-space approach to topological order cannot separate different regions of inhomogeneous samples.

Finally we analyze the present results from the viewpoint of  the modern theory of the insulating state.\cite{rap132,rap_a31} Both \eqs{chern}{trc} look like the imaginary offdiagonal part of a more general tensor; its corresponding symmetric part is real and measures indeed the localization of the electronic ground state in any homogeneous insulator. The key ingredient  is the localization tensor $\lt$, a.k.a. second cumulant moment of the electron distribution (Greek subscripts are Cartesian indices); it has the dimensions of a squared length and its trace provides the gauge-invariant part of the quadratic spread of the Wannier functions, according to the Marzari-Vanderbilt theory.\cite{Marzari97} Notice that localized Wannier functions {\it do not exist} whenever $C \neq 0$;\cite{Thouless84} nonetheless $\lt$ remains well defined and finite in any insulator.\cite{rap127,Thonhauser06}

The direct link between $\lt$ and $C$ has been investigated elsewhere within periodic boundary conditions;\cite{rap_a31,rap127} to see the relationship to  \equ{trc} we write the localization tensor as\cite{rap118,rap132} \[ \lt = \frac{1}{N_{\rm c}} \mbox{tr}_{\rm cell} \{ r_\alpha P r_\beta Q \} , \label{rab}  \] whence 
$ C = 4\pi n_0 \,  \mbox{Im } \langle r_1 r_2 \rangle_{\rm c}  $ ($n_0 = \nc/A_{\rm c}$ is the density). It has been shown\cite{rap_a31,rap132} that \equ{rab} generalizes to finite systems within open boundary conditions, just taking the trace over the whole system and dividing by the total number of electrons. The tradeoff is that $\lt$ becomes then real symmetric, in agreement with the present findings.

In conclusion, we have addressed a system of independent spinless electrons in 2d, whose topological order is classified by means of the archetypical topological invariant: the Chern number $C$. We have found the explicit form of a {\it local} Chern number $\C(\r)$. It is a gauge-invariant microscopic function,  whose macroscopic average coincides with $C$ in crystalline samples. Notably, the boundary conditions (either periodic or open) are irrelevant in the definition of $\C(\r)$. For disordered and/or inhomogeneous samples  the macroscopic average of $\C(\r)$ is a marker which detects the kind of topological order in any  macroscopically homogeneous region: for instance either in a disordered sample, or across an heterojunction. 

At the root of our local description of topological order is the ``nearsightedness'' of the ground-state density matrix. Since this is a very general feature of  insulators, it is possible that any kind of topological order\cite{Moore10,Hasan10}---described by invariants other than $C$---could be addressed via the appropriate local marker.

R. R. acknowledges invaluable discussions with David Vanderbilt about the Chern invariant and more. Work partially supported by the ONR Grant N00014-11-1-0145.
 

\end{document}